# Charge transfer and trapping as origin of a double dip in the transfer characteristics of graphene based field-effect transistors

*Antonio Di Bartolomeo*[1,*], *Filippo Giubileo*[2,1], *Salvatore Santandrea*[1], *Francesco Romeo*[1,2], *Roberta Citro*[1,2], *Thomas Schroeder*[3], *And Grzegorz Lupina*[3]

[1] Dipartimento di Fisica E. R. Caianiello and Centro Interdipartimentale Nanomates, Università degli Studi di Salerno, via Ponte don Melillo, 84084 Fisciano (SA), Italy

[2] CNR-SPIN Salerno, via Ponte don Melillo, 84084 Fisciano (SA), Italy

[3] IHP, Im Technologiepark 25, 15236 Frankfurt (Oder), Germany

**Abstract**

We discuss the origin of an additional dip other than the charge neutrality point observed in transfer characteristics of graphene-based field-effect transistors. The double-dip is proved to arise from charge transfer between graphene and metal electrodes, while charge storage at the graphene/$SiO_2$ interface enhances it. Considering different Fermi energy from the neutrality point along the channel and partial charge pinning at the contacts, we propose a model which explains all features in gate voltage loops.

---

\* To whom correspondence should be addressed. E-mail: dibant@sa.infn.it.

Graphene field effect transistors (GFETs) have attracted substantial interest for applicability to high-speed electronics and spintronics and have been extensively used to investigate electronic transport properties of graphene. In such devices, an electric current is injected/extracted from metallic electrodes (source/drain) through a graphene channel whose conductance is modulated by the electric field from a back- or top-gate. The linear energy dispersion, with zero bandgap and a double-cone shape with intrinsic Fermi level at the vertex, gives symmetric valence and conduction bands; differently from most materials, current modulation by means of a gate in GFETs is possible even without a bandgap, owing the low density of states in two-dimensional graphene[1,2].

Metal/graphene interfaces have been shown to play a significant role in the electrical characteristics of the transistors and various metals (Al, Au, Co, Pd, Pt, Ti, …) have been employed as electrodes. Transfer characteristics of GFETs, i.e. the drain-to-source current vs gate voltage, $I_{DS}$-$V_{GS}$ curves, typically display a symmetric V shape, with a hole dominated conductance (p-branch) at lower $V_{GS}$ and electron type transport at more positive gate voltages (n-branch), the valley corresponding to the charge neutrality (also known as Dirac point) with equal electron and hole concentrations; however, asymmetric[3,4] and/or anomalously-distorted p-branches[5-7] have been reported. The asymmetry between p- and n-branch was initially explained in terms of different cross section of electron/hole scattering from charge impurities[8,9] but more recently the metal/graphene interaction at the contact has been considered as key ingredient[4,10-13]. It has been found in particular that, even in case of weak adhesion, as with Au, the metal electrode cause the Fermi level $E_F$ to shift from the conical point in graphene, resulting in doping of graphene either with electrons or holes; the amount of doping can be deducted from the difference of the metal and graphene workfunctions ($\Phi_M$-$\Phi_G$) and from the potential step ($\Delta V$) due to the metal/graphene chemical interaction ($E_F = \Phi_M - \Phi_G - \Delta V$)[10-13]. Depending on the polarity of carriers in the bulk of the graphene channel, charge transfer between metal and graphene leads to p-p, n-n or p-n junctions in the vicinity of the contacts which can cause asymmetry.

Nouchi et al.[5,6] have studied transfer characteristics in devices with ferromagnetic metal electrodes, reporting anomalously distorted p-branches, with a sort of additional minimum other than the Dirac point. They explain such effect considering charge transfer from graphene to metal leads and assuming that the presence of an oxide layer spontaneously formed at the metal/graphene interface suppresses the charge-density pinning effect, i.e. the modulation of charge density of graphene at the metal electrodes by the gate voltage. A second conductance minimum to the left of the original Dirac point has been also very recently investigated by Chiu et al.[14] for Ti contacted graphene transistors in the high field regime. They showed that the original Dirac point stays unaffected, while the position of a second Dirac point caused by a drain stress depends on the back-gate voltage and they argue that a positive charge is trapped at the graphene/oxide interface in the vicinity of the drain; such charge induces the formation of a p-n junction in drain region and accordingly they suggest a model based on a step-potential to account for the observed double Dirac point.

A double dip in the transfer characteristic has been also discussed by Barraza-Lopez et al.[15] with a first-principles study of the conductance through graphene suspended between Al contacts. They show that the charge transfer at the leads and into the freestanding section gives rise to an electron-hole asymmetry in the conductance; more importantly they suggest that, for sufficiently long junctions, this charge transfer induces two conductance minima at the energies of the two Dirac points of suspended and clamped regions, respectively.

In this paper we present measurements on Cr/Au-contacted long-channel (~10 μm) graphene transistors on Si/SiO$_2$ substrate. We report the observation of hysteresis as well as double dips in the transfer characteristics, that, as far as we know, have never been reported before on GFET with Cr/Au electrodes. Charge trapped in the surrounding dielectric and in particular in silanol groups at the SiO$_2$ surface is at the origin of hysteresis; on the other hand, the gradient of carrier along the channel caused by electron transfer from the graphene to the Au/Cr contacts and the band shift induced by the back-gate voltage and the SiO$_2$-trapped charge are proposed to accounts for the double-dip feature.

We show that p-n junctions are spontaneously formed by charge transfer between graphene and electrodes, and a double Dirac point can be achieved when low-resistivity contacts are fabricated. We further clarify the role of charge stored at the SiO$_2$ interface in the formation of the double dip. Theoretical modeling of experimental data was successfully implemented by taking into account a different doping at contacts with respect to the bulk channel and partial charge pinning at the contacts.

We finally show that a double-dip enhanced hysteresis can conveniently be exploited to build graphene-based memory devices.

Micron-scale graphene flakes were deposited by scotch-tape method on 300 nm thick SiO$_2$ thermally grown on top of a highly p-doped Si substrate. Natural graphite flakes (from NGS Naturgraphit GmbH) were repeatedly cleaved with adhesive tapes and then transferred to SiO$_2$ substrates. The surface of the chip was inspected by optical microscopy to identify suitable few- and monolayer graphene flakes according to the color contrast[16]. Single layer graphene flakes were further confirmed by Raman spectroscopy[17]. Metal contacts of Cr/Au (5nm/150nm, with Cr as adhesion layer) were sputtered after electron beam lithography and structured by lift-off, on selected single-layer graphene flakes. Soon after, some devices were covered by 250 nm thick polymetyl methacrilate (PMMA). PMMA was spin-coated on the whole chip and cross-linked, thus made resistant to acetone etch, by exposure to 30 keV electrons at a dose of $3 \cdot 10^4$ μC/cm$^2$, on the device area.

Figure 1a and 1b show the layout and SEM top view of a typical device before PMMA coverage, respectively. Figure 1c shows the Raman spectrum of the flake used as bulk channel of the GFET, with G and 2D peaks typical of single layer graphene. Single layer graphene in junction with a few layer graphene film (as in figure 1) were preferred in the attempt of minimizing the graphene doping due to interaction with SiO$_2$ substrate[18].

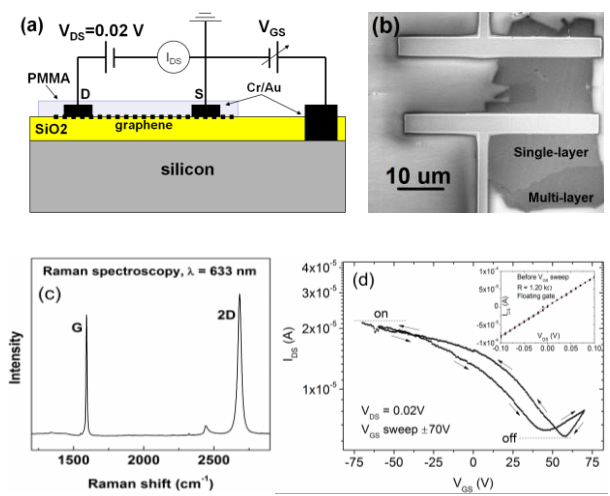

**Figure 1.** (a) Cross section of the device layout and electrical measurement setup. (b) SEM image of a single layer graphene with Cr/Au electrodes (source/drain). Channel width is W ≈ 23 μm and channel length is L ≈ 12 μm. The effective device has been contacted such that it involves only the single-layer part of a multi-layer graphene flake. (c) Raman spectrum showing 2D and G peaks typical of a single layer graphene. (d) Transfer characteristic $I_{DS}$-$V_{GS}$ of the device; the inset shows the source-to-drain current measure at floating gate.

We performed 3-terminal measurements, with the Si substrate as the back-gate and the metal electrodes as the source and drain. All the measurements were performed in air and at room temperature using a HP4140B semiconductor parameter analyzer. Back-gate voltage sweeps, in the interval (-80V, 80V), were performed at constant low drain bias (20 mV). Higher gate voltages were avoided to prevent oxide damage; indeed higher voltage stresses ($V_{GS}$>100V) were often observed to increase gate leakage till oxide breakdown.

Figure 1d shows the transfer characteristic of the device. A minimum $I_{DS}$, i.e. a conductance minimum, corresponding to the charge neutrality point, is observed at $V_{GS}^* \sim 60$ (45) V in an initial reverse (forward) $V_{GS}$ sweep with amplitude of 70 V (figure 1d); the low on/off ratio of about 5 is expected for a graphene flake with low or zero bandgap. A positive charge neutrality point $V_{GS}^*$ indicates that the graphene is unintentionally highly p-doped. We observed this behavior on all fabricated GFETs; indeed, the Dirac point was often located behind the sweeping upper limit of 80 V, especially for devices not covered by PMMA. The formation of weak C-O bonds between graphene and $SiO_2$ has been proven[18,19] to support p-type conductivity in graphene by transfer of charge from the carbon in graphene to the oxygen of the $SiO_2$. This increases the hole concentration in graphene and favors the formation of p-type conductivity at unbiased gate, thus forward shifting the Dirac point.

Moreover, molecules adsorbed on the surface of the channel or at the graphene/$SiO_2$ interface during the fabrication process, consisting mainly of hydrocarbons, carbon dioxide, oxygen and water, have been proven to be a further cause of forward shift of the charge neutrality point. Indeed, for $H_2O$[20], $CO_2$ and $O_2$[21], it has been shown that there is an electron transfer from graphene to the adsorbed molecules, which results in p-doped graphene. Consequently, the use of PMMA as GFET coverage prevents further adsorption and maintains the neutrality point within the swept $V_{GS}$ range. That enabled us to measure and study both the p- and part of the n-branch of the transfer characteristic.

A second important feature observed in the measured curve (figure 1c) is a clear hysteresis. Several recent reports have shown a strong hysteretic behavior in the field effect characteristics of Si/$SiO_2$ supported GFET. In analogy to single-walled carbon-nanotube based field-effect transistors[22,23], gate hysteresis has been attributed mainly to charge trapping in silanol groups (Si-OH) with surface-bound $H_2O$ molecules facilitating the process of charge transfer and trapping[24,25]. Consequently the concentration, distribution and reactivity of the silanol groups of the underlying $SiO_2$ plays a decisive role in the transfer characteristics of a GFET. High concentration of silanol groups makes hydrophilic the $SiO_2$ surface (in general dipolar molecules can easily attach to SiOH), but special treatment can turn this surface hydrophobic; indeed, nearly hysteresis-free GFET have been be achieved on SiO$_2$/Si substrates covered by a thin hydrophobic self-assembled organic layer of HMDS solution (hexamethyldisilazane/acetone 1:1)[26]. Thermal annealing or vacuum pumping can also help reducing hysteresis[14]. Nevertheless, we decided not to apply any treatment or annealing (other than the electrical one) to avoid the risk of introducing damages or stresses to the devices.

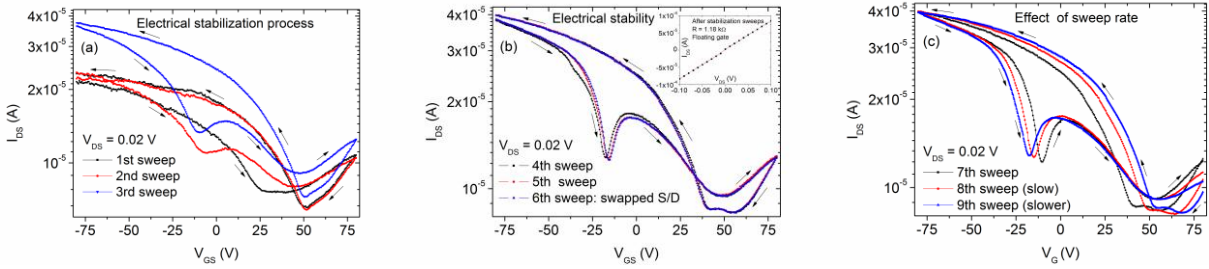

**Figure 2.** Transfer characteristics of the GFET of figure 1 (a) during the stabilization process and (b) after stabilization (inset: I$_{DS}$-V$_{DS}$ for floating gate after stabilization). The stabilization results in higher current and in the appearance of a double dip. (c) Effect of sweep rate.

In figure 2 we report the evolution of transfer characteristics for successive V$_{GS}$ sweeps, acting as electrical annealing. Remarkably, figure 2a, shows that the electrical cycles produce an increase of the current and the appearance of a double Dirac point, i.e. of two conductance minima. Further sweeps demonstrate the stabilization of the device (figure 2b). Swapping drain and source has no effects. Two clear dips appear both in the reverse (where they are closer and less pronounced) and in forward V$_{GS}$ sweep. Figure 2c shows that slower sweeping rate, which favors charge injection and trapping at the SiO$_2$ surface, widens the hysteresis loop.

The electrical stabilization is the result of current self-annealing, mainly acting at the graphene-electrode interface by reducing/stabilizing the contact resistance, while PMMA coverage prevents removal of contaminants from the graphene surface[7] that may induce changes in the bulk channel. The contact resistivity[27] , $\rho_C$ = RW~5 kΩμm (W is the width of the channel), is in low side of the range usually reported for Cr/Au contacted GFETs (2·10$^3$ Ωμm≤ $\rho_C$ ≤10$^6$ Ωμm) [28]; $\rho_C$ is estimated at V$_{GS}$= -80 V when the source-to-drain resistance is dominated by the contacts, the graphene bulk channel being at its maximum conductance. A further confirmation of the good contacts stems from the calculation of the mobility[29]. Despite the top coverage, which may affect the mobility[30] for the participation of π-orbitals to the van der Waals bonds to the PMMA, holes mobility of ~3000

cm$^2$/Vs, thus exceeding the universal mobility of silicon, are achieved in our devices; this value is consistent with measurements performed on devices of comparable geometry where metals with better chemisorption to graphene have been used[31,32]. The mobility µ has been here estimated from the simple formula $\mu = \frac{dI_{DS}}{dV_{GS}} \frac{L}{W} \frac{d}{\varepsilon} \frac{1}{V_{DS}}$ where L and W are the channel length and width, d the SiO$_2$ thickness, ε its dielectric constant, $dI_{DS}/dV_{GS}$ the transconductance calculated from the linear region of the $I_{DS}$-$V_{GS}$ curves at the left side of the Dirac point.

We notice that the double Dirac point is obtained while keeping the drain bias very low (20 mV), suggesting that it could not be only the result of charge trapped in the vicinity of the drain at the graphene/SiO$_2$ interface during the drain stress[14].

In the following we suggest that charge transfer between graphene and metal contacts and charge trapping at SiO$_2$/graphene interface can fully explain the behavior of the $I_{DS}$-$V_{GS}$ loop and account for the double Dirac point appearance. We suggest that the double dip is the result of a different Fermi level alignment within the graphene double-cone at the contacts with respect to the bulk channel; the Fermi level within the band diagrams is shifted by the back-gate voltage and is influenced by the charge trapped at the SiO$_2$/graphene interface.

Due to different workfunctions (4.6eV for Cr, 5.1eV for Au and 4.5eV for graphene[11]) electrons transfer from the graphene to the metal electrodes, thus forming a doping gradient from the contacts to the bulk channel. Underneath and close to the electrodes, the graphene is more p-doped than in the channel[10]. The doping of the graphene by the contacts is not limited only underneath the metal electrodes but extends for 0.2–0.3 µm[33] or longer[34] in the inner channel, since the graphene, having zero density of states at the Dirac point, is not able to absorb all the transferred holes at the interface.

While charge density pinning (i.e. gate uncontrollability of charge density at the metal contacts) could occur at Au/graphene contacts[4,12,33], reactive materials have been proven to lead to charge depinning[6], especially when an oxide layer is formed at the metal interface. We assume here partial charge pinning with charge at the contacts controlled by the back-gate up to a certain limit, over which the charge cannot be increased. This is a reasonable assumption because the low contact resistance makes the potential of the graphene at the contacts anchored to the bias of source and drain; consequently, the field of the back-gate is expected to affect mainly the carrier concentration in the bulk graphene channel that at the contacts. We will show that the assumption of partial pinning leads to a good fit of the experimental $I_{DS}$-$V_{GS}$ curves.

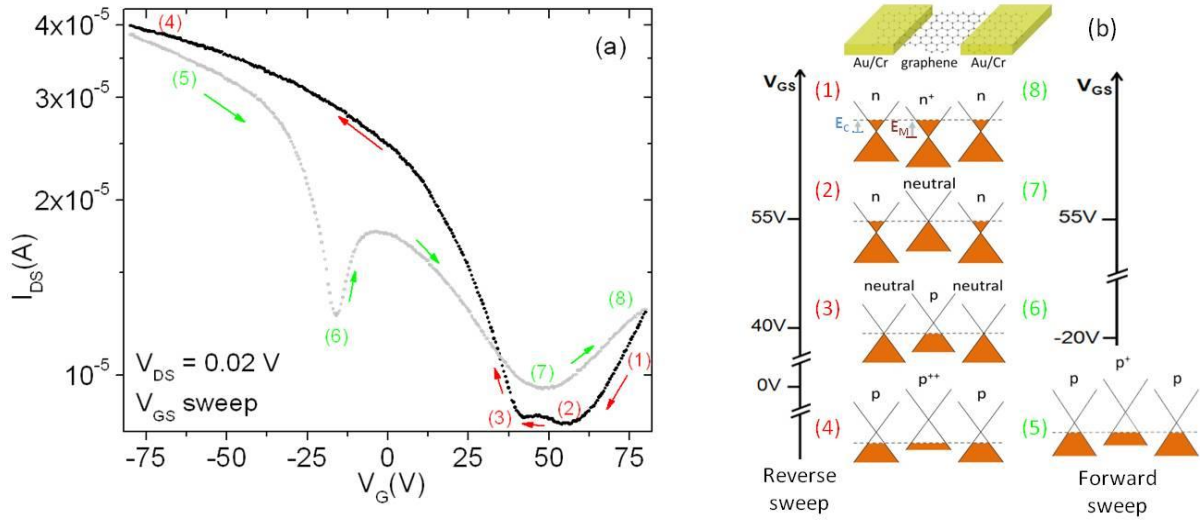

**Figure 3.** (a) Transfer characteristic of a GFET with hysteresis and double dip. (b) Band diagrams of graphene between source and drain and position of the Fermi level at different $V_{GS}$ for different gate voltages. For floating gate, the double cone, close to the contacts, is shifted upward with respect to the one in the bulk channel to account for the p-type graphene due to transfer of electrons from graphene to the Cr/Au leads.

The effect of charge transfer at the contacts can be taken into account by shifting the energy band diagram (double cone) of the graphene upward with respect to that in the bulk channel for unbiased $V_{GS}$ (figure 3). The application of the back-gate voltage moves the Fermi level with respect to the double-cone determining different conduction regions between source and drain. At $V_{GS} \geq 80V$, n-type conduction takes place everywhere thus giving a high conductance (point 1 in figure 3a). The source-to-drain channel appears as a n/n$^+$/n structure, where the notation corresponds to the graphene doping of source/channel/drain regions. While lowering $V_{GS}$ (reverse sweep) electrons move away through contacts, so a charge neutrality condition, i.e a first Dirac point, is reached in the channel before than at the contact regions (where graphene stay longer n-doped). Consequently, the n/n$^+$/n junction gradually evolves into a n/neutral/n structure. The resistance associated with a neutral/n junction is larger than that of an equivalent n/n$^+$ (or sharp n/p) junction as can be easily understood within a diffusive carrier transport model, where the resistance is simply obtained by integrating the local resistivity along the length of the junction. Therefore this condition results in a conductance minimum which corresponds to the first valley (point 2) in the transfer characteristic.

A further decrease in $V_{GS}$ generates a p-type conduction in the inner channel, giving rise to a $n^-/p/n^-$ structure, with slightly increased conductance. When the graphene at the contacts reaches the charge neutrality point (neutral/p/neutral junction) a second Dirac point (point 3) is observed. The separation $\Delta V_{GS} \sim 10$ V of these two dips is related to the initial carrier concentration difference between the channel and contact regions, i.e. to $\Delta E_{MC} = E_M - E_C$ at the beginning of the sweep (here $E_M$ and $E_C$ are the Fermi levels of graphene with respect to the local neutrality point in the channel and at the metal contacts, respectively). Going to more negative $V_{GS}$ a low-resistance p channel ($p/p^{++}/p$ structure) is formed; the pinning of charge at the contacts is in this case responsible for the quasi-saturation observed in the transfer characteristic at high negative $V_{GS}$.

On the other hand, the high negative voltage applied during the reverse sweep makes positive charge to be stored at the $SiO_2$/graphene interface, which, as already said, is at the origin of the leftward shift of the transfer characteristic during the following forward sweep, i.e. of the hysteresis. This trapped charge acts as reduction of the overall p-doping and can be taken into account with a down-shift of the graphene bands (configuration 5). During the forward sweep the p-doping of the graphene is neutralized by attracting electrons from the contacts and a neutral condition is soon reached at the contacts, thus originating a first Dirac point (point 6 at $V_{GS} \sim -18V$ in figure 3a). Further increase of $V_{GS}$ creates a second dip (at $V_{GS} \sim 48V$) when neutrality is reached in the bulk channel (point 7). The second Dirac point happens at a $V_{GS}$ value slightly below the one observed during the reverse sweep as effect of the charge stored at the $SiO_2$/graphene interface; finally a low-conductance $n/n^+/n$ structure (point 8) is formed again. The $\Delta V_{GS} \sim 65V$ separation between the two dips in the forward sweep is increased with respect to the previous $\Delta V_{GS} \sim 10V$ since the downward band shift created by the $SiO_2$ trapped charge and the injection of electrons from the contacts greatly favors the appearance of a Dirac point in the contact regions. The position of this point, being related to the $SiO_2$ trapped charge, depends also on the maximum negative voltage applied during the previous reverse sweep. Indeed, inset of figure 4 shows that the Dirac point appears at higher voltages ($V_{GS} \sim 40V$) when we start at $V_{GS} = 0V$ after letting the device release the $SiO_2$ trapped charge, by a room temperature annealing of few days.

Very importantly, our model predicts that, whatever is the choice of the metal leads, the extra Dirac point appears always in the p-branch of the transfer characteristic. This is in agreement with ref. 6 and 14 where a second dip was not found when stressing Ti or Co-contacted GFETs under n-type transport conditions; a peculiarity that has been explained[14] through the unlikely assumption that positive charges are much easier to be injected into the $SiO_2$ trap centers than electrons.

To check the proposed phenomenological model we have performed a numerical simulation exploiting the Boltzmann transport equation in describing the diffusive dynamics of the device. Assuming that the charge stored at the $SiO_2$/graphene interface is uniformly distributed across the channel and that its density $\rho_{it}$ is at most a second power on $V_{GS}$[22,23], the total carrier density $\rho$ along the graphene channel is a linear function of $\rho_{it}$ and, with good approximation, a quadratic function of the Fermi level $E_F$, which in turn is proportional to $V_{GS}$. The dependence of $\rho$ on $E_F$ can be extracted by the energy dependence of the density of states ($D_\lambda(E) \propto |E|$) observing that the integral $\rho \propto \rho_{it} + \sum_\lambda \lambda \int_0^\infty D_\lambda(E) f_\lambda(E) dE$ ($f_\lambda(E)$ is the Fermi function and $\lambda = \pm 1$ for electrons and holes, respectively) roughly shows a quadratic behaviour on $E_F$ whose temperature dependence only modifies a constant prefactor. Accordingly we have assumed an overall quadratic dependence of $\rho$ on $V_{GS}$. The value of $\rho$ so obtained has been used to calculate the source-to-drain conductance G: $G \propto \frac{4e^2}{h} \frac{W}{L} \frac{\hbar v_F}{k_B T} \rho(V_{GS})$, with $v_F \approx 10^6 \, m/s$ is the Fermi velocity, T the temperature[1]. To take into account the mismatch of the Fermi level at the contacts and in the bulk channel, we have divided the channel in 3 regions and taken $E_F = E_C$ at the two contact regions and $E_F = E_M$ in the bulk channel (with $|E_M| > |E_C|$ at $V_{GS} = 0$). The conductances of the three regions have been combined in series; we have also included a parasitic resistance originating from the contacts and leads.

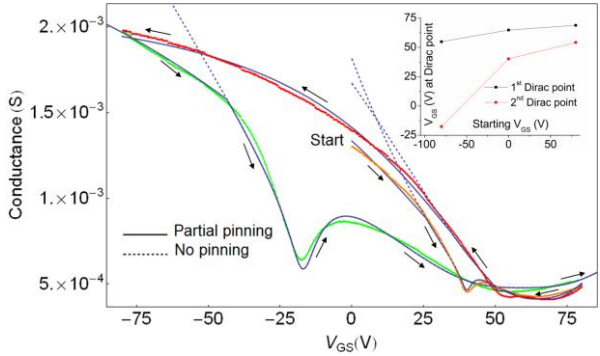

**Figure 4.** Conductance vs gate voltage for the device with no charge initially stored on $SiO_2$ (orange, red and green curves are experimental data). The dashed and full black curves represent the fit of our model whithout and with partial charge pinning at the contacts. The inset shows the $V_{GS}$ positions of the two Dirac points as a function of the starting gate voltage.

Figure 4 shows the predicted G-$V_{GS}$ behavior superposed to the experimental data on an entire loop starting at $V_{GS}$ = 0. We have calculated the conductance G both without (dashed lines) and with (full line) partial charge pinning at the contacts (i.e upper limit for $|E_c|$). As can be seen in figure 4, the agreement of the proposed model with experimental data is satisfactory and the simulation clearly supports our hypothesis of partial charge pinning. An additional element pointing towards our interpretation is obtained from the consideration that the pinned value of $|E_c|$ (as extracted from the experimental data to reproduce the quasi saturation behavior) asymptotically reaches the difference between the gold and graphene workfunction (i.e. $\Phi_{Au} - \Phi_G \approx 0.6 eV$).

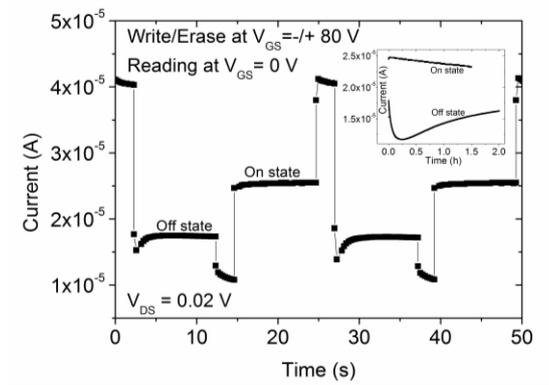

**Figure 5.** GFET used as memory device. A $V_{GS}$ = 80V pulse set the device to its *off* state, while a $V_{GS}$ = -80V pulse switches it to the *on* state. Reading is performed at $V_{GS}$ = 0V. On and off state constitute the two logic levels of a memory device. Insert in figure shows that those states stay separated for a time at least of two hours.

We highlight that the presence of a double dip results in an effective widening of the hysteretic loop. Hysteresis observed in carbon-nanotubes based transistors has been be exploited to build memory devices[23,35]. Similarly, here, the two values of the current at a given gate voltage, can be considered as the two logic levels (on and off state) of a memory device; positive or negative $V_{GS}$ pulses can be used to switch between these two states to implement the write and erase operations. Figure 5 is a proof of the concept, where reading is performed at $V_{GS}$ = 0V and ±80 V pulses are used for write and erase. The programming pulses, in our case, are too big for practical applications; nevertheless, with proper treating of the $SiO_2$/Si or by ionic screening[36] it could be possible to obtain transfer characteristics with the deeper dip around $V_{GS}$ = 0 and operate the device with lower pulses. For practical technological applications, the same device could be implemented with a top-gate.

In conclusion, we have clarified the nature of a double dip often observed in the p-branch of the transfer characteristic of a GFET. We have shown that it is related to charge transfer between graphene and metal contact and that it is enhanced by the hysteresis provoked by charge storage at the graphene/$SiO_2$ interface. Elucidating the origin of such anomaly is of technological importance since the observed distortion indicate a deterioration of gate-voltage response of the device. i.e. a decrease in field modulation of channel conductance. Although a possible drawback in circuitry applications, we have also suggested that such feature can be conveniently exploited to develop graphene based memory devices.

**Acknowledgments**

We thank Prof. P. Barbara, Dr. Y. Yang, Dr. A.K. Boyd and Dr. M. Rinzan for the logistic and technical support and the fruitful discussions during the time spent for the fabrication and characterization of the devices at the Georgetown Nanoscience and Microtechnology Laboratory (GNuLab) of the Georgetown University, in Washington DC.